\documentclass[a4paper,11pt]{article}
\usepackage{jcappub} % for details on the use of the package, please see the JINST-author-manual
\usepackage{lineno}
\usepackage[margin=5pt,caption=false]{subfig}
\usepackage{slantsc}
\usepackage{comment}
\usepackage{threeparttable}
\usepackage{booktabs}
\usepackage[figuresright]{rotating}
\usepackage{textcase}
\usepackage{multirow}
\newcommand{\be}{\begin{equation}}
\newcommand{\ee}{\end{equation}}
\newcommand{\bea}{\begin{eqnarray}}
\newcommand{\eea}{\end{eqnarray}}
\newcommand{\hunit}{$\rm{km \ s^{-1} \ Mpc^{-1}}$}
\newcommand{\lcdm}{$\Lambda$CDM}
\newcommand{\pcdm}{$\phi$CDM}
\usepackage{array}

\newcommand{\hiig}{H\,\textsc{ii}G}
\newcommand{\hii}{H\,\textsc{ii}}
\newcommand{\Om}{\Omega_{m0}}
\newcommand{\Ok}{\Omega_{k0}}
\newcommand{\om}{$\Omega_{m0}$}
\newcommand{\ok}{$\Omega_{k0}$}
\newcommand{\wx}{$w_{\rm X}$}
\newcommand{\wX}{w_{\rm X}}

\newcommand{\mii}{Mg\,\textsc{ii}}

\newcommand{\civ}{C\,\textsc{iv}}

\newcommand{\obh}{\Omega_{b}h^2}
\newcommand{\och}{\Omega_{c}h^2}
\newcommand{\onh}{\Omega_{\nu}h^2}
\newcommand{\obhs}{$\Omega_{b}h^2$}
\newcommand{\ochs}{$\Omega_{c}h^2$}

\usepackage[T1]{fontenc}
\usepackage{scalerel}
\usepackage{tikz}
\usetikzlibrary{svg.path}

\definecolor{orcidlogocol}{HTML}{A6CE39}
\tikzset{
  orcidlogo/.pic={
    \fill[orcidlogocol] svg{M256,128c0,70.7-57.3,128-128,128C57.3,256,0,198.7,0,128C0,57.3,57.3,0,128,0C198.7,0,256,57.3,256,128z};
    \fill[white] svg{M86.3,186.2H70.9V79.1h15.4v48.4V186.2z}
                 svg{M108.9,79.1h41.6c39.6,0,57,28.3,57,53.6c0,27.5-21.5,53.6-56.8,53.6h-41.8V79.1z M124.3,172.4h24.5c34.9,0,42.9-26.5,42.9-39.7c0-21.5-13.7-39.7-43.7-39.7h-23.7V172.4z}
                 svg{M88.7,56.8c0,5.5-4.5,10.1-10.1,10.1c-5.6,0-10.1-4.6-10.1-10.1c0-5.6,4.5-10.1,10.1-10.1C84.2,46.7,88.7,51.3,88.7,56.8z};
  }
}

\newcommand\orcidicon[1]{\href{https://orcid.org/#1}{\mbox{\scalerel*{
\begin{tikzpicture}[yscale=-1,transform shape]
\pic{orcidlogo};
\end{tikzpicture}
}{|}}}}
\usepackage{amsmath,bm}
\usepackage{amssymb}
\usepackage{fnpct}

\arxivnumber{2404.08697} % Only if you have one
\title{\boldmath Testing the standardizability of, and deriving cosmological constraints from, a new Amati-correlated gamma-ray burst data compilation}

\author[a,b]{Shulei Cao$^{\orcidicon{0000-0003-2421-7071}}$,}
\author[a]{Bharat Ratra$^{\orcidicon{0000-0002-7307-0726}}$}
\affiliation[a]{Department of Physics, Kansas State University,\\
116 Cardwell Hall, Manhattan, KS 66506, USA}
\affiliation[b]{Department of Physics, Southern Methodist University,\\
3215 Daniel Ave Fondren Science Building, Dallas, TX 75205, USA}

\emailAdd{shuleic@mail.smu.edu, ratra@phys.ksu.edu}

\abstract{By using gamma-ray burst (GRB) data to simultaneously constrain Amati correlation parameters and cosmological parameters in six spatially flat and nonflat dark energy cosmological models, we show that an updated 220 GRB version of the Jia et al.\ \cite{Jiaetal2022} GRB data compilation are standardizable through the Amati correlation and so can be used for cosmological analyses. However, the resulting GRB data constraints on the current value of the nonrelativistic matter density parameter, $\Omega_{m0}$, are in $>2\sigma$ tension with those from a joint analysis of better-established Hubble parameter [$H(z)$] and baryon acoustic oscillation (BAO) data for most of the cosmological models we consider, indicating that these GRB data cannot be jointly used with better-established $H(z)$ + BAO data to constrain cosmological parameters.}

\begin{document}
\maketitle
\flushbottom

\section{Introduction}
\label{sec:intro}

A number of independent observations indicate that the Universe is currently undergoing accelerated cosmological expansion. The prevailing explanation for this phenomenon is dark energy, a hypothetical substance characterized by negative pressure. Among dark energy cosmological models, the most popular spatially flat \lcdm\ model, \cite{peeb84}, assumes dark energy is a cosmological constant, $\Lambda$, contributing approximately 70\% of the total energy budget of the current Universe. However, some recent observations hint at potential deviations from this model (see e.g.\ Ref.\ \citep{PerivolaropoulosSkara2021,Morescoetal2022,Abdallaetal2022,Hu:2023jqc}), prompting exploration of alternate cosmological models accommodating non-zero spatial curvature and dynamical dark energy. In our analysis here we use some of these alternate models to check the standardizability of gamma-ray burst (GRB) data.

At present, better-established cosmological observations predominantly explore either the lower-redshift Universe, at $z < 2.3$, \cite{CaoRatra2023}, or the Universe at $z \sim 1100$, \cite{planck2018b}. In the intermediate redshift range GRB data extending up to $z\sim8.2$ serve as potential cosmological probes, through correlations the GRBs obey that allow some of them to be standardizable \citep{Dirirsa2019, KhadkaRatra2020c, Khadkaetal_2021b, ZhaoXia2022, CaoDainottiRatra2022b, DainottiNielson2022, Liuetal2022b, GovindarajDesai2022, Liangetal2022, Singhetal2024, Kumaretal2023, Lietal2023, Muetal2023, Lietal2023b, Xieetal2023, Zhangetal2023, Favaleetal2024}. We have analyzed GRB data by simultaneously constraining correlation parameters and cosmological parameters across various cosmological models. This approach is crucial to circumvent the circularity problem and to ascertain whether the correlation is independent of cosmological model, thus potentially rendering it standardizable \citep{KhadkaRatra2020c, Caoetal_2021}. 

Our analysis reveals that a subset of 118 Amati-correlated (A118) GRBs, characterized by lower intrinsic dispersion, is indeed independent of the assumed cosmological model, making these data standardizable and suitable for cosmological investigations \citep{Khadkaetal_2021b, LuongoMuccino2021, CaoKhadkaRatra2021, CaoDainottiRatra2022, Liuetal2022, Liuetal2022b, Zhangetal2023, WangLiLiang2024, NongLiang2024}.\footnote{Our method of analyzing GRB data differs from other cosmological model-independent methods which first calibrate GRB data correlations by either SNe Ia (see e.g., Ref.\ \cite{Liangetal2008,LiangWuZhang2010,LiangXuZhu2011,Wei2010,Liu_Wei_2015,Wang_2016,Demianski_2017a}) or $H(z)$ data (see e.g., Ref.\ \cite{Amati2019,LuongoMuccino2021,Montiel2021,Lietal2023,WangLiang2024}) and then constrain cosmological parameters with calibrated correlations. Typically the calibration is done using a Gaussian process summary of the calibration data, so while these results do not depend on an assumed cosmological model they do depend on the assumed Gaussian processes kernel. Additionally in a number of cases GRB data only cosmological constraints are not consistent with those derived from the calibration data and it is then incorrect to calibrate the GRB dataset in these cases.} Cosmological constraints from the A118 GRB dataset are consistent with, but weaker than, those from better-established data, so A118 data can be jointly used with these better-established data to constrain cosmological parameters, with the resulting cosmological constraints being slightly more restrictive when A118 GRB data are included in the mix. However, the A118 dataset is small and it would be good to have a bigger standardizable GRB dataset, both to presently more restrictively constrain cosmological parameters as well as to help better understand such GRB data in preparation for large, better-quality, GRB compilations that should soon become available.\footnote{The SVOM space mission \cite{Atteiaetal2023} is scheduled to be launched later this year and THESEUS \cite{Amatietal2021} is an example of a possible future GRB space mission.} 

Recently Ref.\ \cite{Jiaetal2022} have collected a larger sample of 221 Amati-correlated GRBs, nearly twice as large as the A118 GRB sample. Here we study an updated version of the  Ref.\ \cite{Jiaetal2022} sample, to determine whether it is standardizable and also study whether these GRB data cosmological constraints are consistent with those from better established data, such as baryon acoustic oscillation (BAO) observations and expansion rate or Hubble parameter [$H(z)$] measurements. 

Other similar emerging cosmological probes include reverberation-measured \mii\ and \civ\ quasar (QSO) measurements reaching to $z\sim3.4$, \citep{Czernyetal2021, Zajaceketal2021, Khadkaetal_2021a, Khadkaetal2021c, Khadka:2022ooh, Cao:2022pdv, Czerny:2022xfj, Caoetal2024}, which our technique, \citep{KhadkaRatra2020c, Caoetal_2021}, shows are standardizable. Bright \hii\ starburst galaxy (\hiig) data reaching to $z\sim2.5$, \cite{Siegel_2005, Mania_2012, Chavez_2014, Chavez_2016, G-M_2019, GM2021, CaoRyanRatra2020, CaoRyanRatra2021, CaoRyanRatra2022, Johnsonetal2022, CaoRatra2022} might eventually become a standardizable probe. However, our technique, \citep{KhadkaRatra2020c, Caoetal_2021}, shows that in the most recent \hiig\ compilation, \cite{GM2021}, both low-$z$ and high-$z$ \hii\ starburst galaxy compilations can individually be standardized, but they obey distinctly different correlations and so cannot be jointly standardized, challenging assumptions made in previous analyses of these data, \cite{CaoRatra2024}. QSO X-ray and UV flux observations, reaching to $z \sim 7.5$, have been explored as a potential cosmological probe utilizing a similar correlation \citep{RisalitiLusso2015, RisalitiLusso2019, KhadkaRatra2020a, Yangetal2020, KhadkaRatra2020b, Lussoetal2020, KhadkaRatra2021, KhadkaRatra2022, Rezaeietal2022, Luongoetal2021, DainottiBardiacchi2022}. However, it has been found that these QSOs lack standardizability, as the correlation is neither independent of cosmological model nor redshift independent, \citep{KhadkaRatra2021, KhadkaRatra2022}, in the latest QSO flux compilation, \cite{Lussoetal2020}. Therefore, these QSO data cannot be utilized for cosmological purposes, \citep{KhadkaRatra2021, KhadkaRatra2022, Petrosian:2022tlp, Khadka:2022aeg, Zajaceketal2024, Wangetal2024}.

As discussed below, we update the 221 GRB dataset of Ref.\ \cite{Jiaetal2022}\footnote{Ref.\ \cite{Xieetal2023} used this 221 GRB dataset to calibrate the Amati correlation by using a Gaussian process summary of Pantheon+ SNe Ia data that does not depend on an assumed cosmological model but does depend on the assumed Gaussian process kernel function. Other analyses of recent Fermi GRB data can be found by Refs.\ \cite{Montiel2021,WangLiang2024}.} by discarding one of the GRBs and updating measurements for eight other GRBs.\footnote{We note that these J220 GRBs significantly differ from the A220 sample of Ref.\ \cite{Khadkaetal_2021b}, which is a combination of the A118 and A102 samples compiled in Ref.\ \cite{Khadkaetal_2021b}. There are 100 common GRBs between these J220 GRBs and A118, and there are 40 common GRBs between these J220 GRBs and A102, so 140 common GRBs between these J220 GRBs and A220; there is thus a 36\% difference between these J220 GRBs and A220 data.} We show, for the first time, that this updated J220 GRB dataset is standardizable through the Amati correlation. However, the constraints on the present value of the nonrelativistic matter density parameter \om\ are mostly in $>2\sigma$ tension compared with those from better-established $H(z)$ + BAO data in four of the six cosmological models we study. This suggests that the updated 220 GRB dataset of Ref.\ \cite{Jiaetal2022} cannot be used to constrain cosmological parameters, hence the A118 GRB dataset \citep{Khadkaetal_2021b, LuongoMuccino2021, CaoKhadkaRatra2021, CaoDainottiRatra2022, Liuetal2022} still is the largest GRB compilation suitable for cosmological purposes. 

The structure of this paper is as follows. In Section \ref{sec:model} we provide a brief overview of the cosmological models employed in our analyses. Section \ref{sec:data} introduces the datasets used while Section \ref{sec:analysis} outlines our methodology. Our main findings are presented in Section \ref{sec:results}, followed by concluding remarks in Section \ref{sec:conclusion}.

\section{Cosmological models}
\label{sec:model}

Some previous investigations (see Refs.\ \cite{Khadkaetal_2021b, LuongoMuccino2021, CaoKhadkaRatra2021, CaoDainottiRatra2022, Liuetal2022} and references therein) have used GRB data that obey the Amati or $E_{\rm p}-E_{\rm iso}$ correlation, \cite{Amatietal2002}, between the GRB's peak photon energy in the cosmological rest frame and the isotropic energy related to the measured GRB bolometric flux, see discussion in Section \ref{sec:data} below. Our aim here is to study the GRB dataset of Ref.\ \cite{Jiaetal2022}. We undertake this analysis within the framework of six relativistic dark energy cosmological models, encompassing both spatially flat and nonflat scenarios.\footnote{For recent studies on spatial curvature constraints, refer to Refs.\ \cite{Oobaetal2018b, ParkRatra2019b, DiValentinoetal2021a, ArjonaNesseris2021, Dhawanetal2021, Renzietal2022, Gengetal2022, MukherjeeBanerjee2022, Glanvilleetal2022, Wuetal2023, deCruzPerezetal2023, DahiyaJain2022, Stevensetal2023, Favaleetal2023, Qietal2023, deCruzPerez2024}, among others.} Our objective is to examine, by simultaneously constraining cosmological model parameters and $E_{\rm p}-E_{\rm iso}$ correlation parameters,  whether these GRB data conform to the $E_{\rm p}-E_{\rm iso}$ correlation in a manner independent of cosmological model and so are potentially standardizable. We are also interested in deriving cosmological parameter constraints using these GRB data to determine whether or not the GRB data alone cosmological constraints are consistent with those derived using better-established data. For these purposes we need $H(z)$ in each cosmological model. $H(z)$ depends on redshift $z$ and the cosmological parameters of the model under study. It is defined by the first Friedmann equation, derived within the framework of general relativity in the Friedmann-Lema\^{i}tre-Robertson-Walker metric.

We consider models with one massive and two massless neutrino species. With an effective number of relativistic neutrino species $N_{\rm eff} = 3.046$ and a total neutrino mass $\sum m_{\nu}=0.06$ eV, the present value of the nonrelativistic neutrino physical energy density parameter is computed as $\onh=\sum m_{\nu}/(93.14\ \rm eV)$, where $h$ is the Hubble constant ($H_0$) scaled by 100 \hunit. Accordingly, $\Om = (\onh + \obh + \och)/{h^2}$, where \obhs\ and \ochs\ denote the present values of the baryonic and cold dark matter physical energy density parameters, respectively. Given our focus on late-time measurements, we ignore the very small contribution from photons to the late-time cosmological energy budget.

We investigate both \lcdm\ models and XCDM parametrizations, which extend the \lcdm\ framework by allowing temporal (but not spatial) variability in the (fluid) dark energy density but with constant dark energy equation of state parameter $w_{\rm DE}=p_{\rm DE}/\rho_{\rm DE}$, where $p_{\rm DE}$ and $\rho_{\rm DE}$ are the dark energy fluid pressure and energy density, respectively. While \lcdm\ models correspond to $w_{\rm DE} = -1$, XCDM parametrizations allow other values of $w_{\rm DE}$. In these models the governing Friedmann equation takes the form
\be
\label{eq:HzLX}
% \resizebox{0.475\textwidth}{!}{%
% $
H(z) = H_0\sqrt{\Om\left(1 + z\right)^3 + \Ok\left(1 + z\right)^2 + \Omega_{\rm DE0}\left(1+z\right)^{1+w_{\rm DE}}}.
% $%
% }
\ee
Here \ok\ represents the current value of the spatial curvature energy density parameter and $\Omega_{\rm DE0} = 1 - \Om - \Ok$ is the current value of the dark energy density parameter. In the \lcdm\ model dark energy is the cosmological constant $\Lambda$, i.e., $\Omega_{\rm DE0} =\Omega_{\Lambda}$. In the XCDM parametrization dark energy is an X-fluid with a dynamical dark energy equation of state parameter, resulting in $\Omega_{\rm DE0} =\Omega_{\rm X0}$. Since GRB data cannot constrain $H_0$ and $\Omega_{b}$, we fix $H_0=70$ \hunit\ and $\Omega_{b}=0.05$ in the GRB analyses. $\{\Om, \Ok\}$ for \lcdm\ and $\{\Om, \wX, \Ok\}$ for XCDM are the free parameters we constrain with GRB data. As for $H(z)$ + BAO data, $\{H_0, \obh\!, \och\!, \Ok\}$ for \lcdm\ and $\{H_0, \obh\!, \och\!, \wX, \Ok\}$ for XCDM are the free parameters constrained. In spatially flat models $\Ok=0$.

We also explore \pcdm\ models (see e.g. Ref.\ \cite{peebrat88,ratpeeb88,pavlov13})\footnote{For recent cosmological constraints on the \pcdm\ models, see Refs.\ \cite{ooba_etal_2018b, ooba_etal_2019, park_ratra_2018, park_ratra_2019b, park_ratra_2020, Singhetal2019, UrenaLopezRoy2020, SinhaBanerjee2021, deCruzetal2021, Xuetal2022, Jesusetal2022, Adiletal2023, Dongetal2023, VanRaamsdonkWaddell2024, Avsajanishvilietal2024} and references therein.}, where a dynamical scalar field $\phi$ is the dark energy. The Friedmann equation for this model is
\be
\label{eq:Hzp}
H(z) = H_0\sqrt{\Om\left(1 + z\right)^3 + \Ok\left(1 + z\right)^2 + \Omega_{\phi}(z,\alpha)}.
\ee
Here the scalar field dynamical dark energy density parameter $\Omega_{\phi}(z,\alpha)$ is computed numerically by simultaneously solving the Friedmann equation \eqref{eq:Hzp} and the equation of motion of the scalar field \be
\label{em}
\ddot{\phi}+3H\dot{\phi}+V'(\phi)=0,
\ee
where $V(\phi)$ is the potential energy density
\be
\label{PE}
V(\phi)=\frac{1}{2}\kappa m_p^2\phi^{-\alpha}.
\ee
In these equations, $m_p$ is the Planck mass and $\alpha$ is a positive constant to be constrained; note that when $\alpha=0$ the \pcdm\ model is the \lcdm\ model. The constant coefficient $\kappa$ can be determined by the shooting method in the cosmic linear anisotropy solving system (\textsc{class}) code, \cite{class}. In the \pcdm\ model, for GRB data $\{\Om, \alpha, \Ok\}$ are to be constrained, while for $H(z)$ + BAO data $\{H_0, \obh\!, \och\!, \alpha, \Ok\}$ are to be constrained. In the flat \pcdm\ model $\Ok=0$.

\section{Data}
\label{sec:data}

In this paper we test whether the updated GRB J220 data set of Ref.\ \cite{Jiaetal2022}\footnote{Note that the updated J220 GRB dataset of Ref.\ \cite{Jiaetal2022} and our paper differs from the GRB A220 dataset of Ref.\ \cite{Khadkaetal_2021b}.} is standardizable through the Amati ($E_{\rm p}-E_{\rm iso}$) correlation, by using these data to simultaneously constrain the Amati correlation parameters and cosmological parameters in half a dozen cosmological models. These GRB data, as well as A118 and $H(z)$ + BAO data used for comparison purposes, are summarized below.

\begin{itemize}

\item[]{\it J220 and A118 GRB sample.} The J220 sample, which is mostly taken from Table 1 of Ref.\ \cite{Jiaetal2022}, consists of 220 long GRBs and spans a wide redshift range, from $0.034$ to $8.2$. Note, however, that our J220 sample differs from the 221 GRBs of Ref.\ \cite{Jiaetal2022} by removing GRB 020127 (which has an unreliable redshift, \cite{Khadkaetal_2021b}) and by updating 8 GRBs (080916C, 090323, 090328, 090424, 090902B, 091020, 091127, 130427A) to values in \cite{Khadkaetal_2021b}. The bolometric fluence $S_{\rm bolo}$ and the rest-frame isotropic radiated energy $E_{\rm iso}$ of these updated GRBs are computed for the energy range $1-10^4$ keV using Eqs.\ (1) and (2) of Ref.\ \cite{Dirirsa2019} with different spectral fitting models than the Band model. We also reverted $E_{\rm iso}$ in units of erg, reported in Table 1 of Ref.\ \cite{Jiaetal2022}, back to $S_{\rm bolo}$, where $E_{\rm iso}=4\pi D_L^2S_{\rm bolo}/(1+z)$ and $S_{\rm bolo}$ is computed in the standard rest-frame energy band $1-10^4$ keV and in units of $\text{erg\,cm}^{-2}$, where we have used the flat \lcdm\ model with $\Om=0.315$ and $H_0=67.4$ \hunit\ to compute the luminosity distance $D_L(z)$
\begin{equation}
  \label{eq:DL}
% \resizebox{0.475\textwidth}{!}{%
    % $
    D_L(z) = 
    \begin{cases}
    \frac{c(1+z)}{H_0\sqrt{\Ok}}\sinh\left[\frac{\sqrt{\Ok}H_0}{c}D_C(z)\right] & \text{if}\ \Ok > 0, \\
    \vspace{1mm}
    (1+z)D_C(z) & \text{if}\ \Ok = 0,\\
    \vspace{1mm}
    \frac{c(1+z)}{H_0\sqrt{|\Ok|}}\sin\left[\frac{H_0\sqrt{|\Ok|}}{c}D_C(z)\right] & \text{if}\ \Ok < 0,
    \end{cases}
    % $%
    % }
\end{equation}
with $D_C(z)$ being the comoving distance
\begin{equation}
\label{eq:gz}
   D_C(z) = c\int^z_0 \frac{dz'}{H(z')},
\end{equation}
and $c$ is the speed of light.

The Amati correlation \citep{Amatietal2002, Amati2008, Amati2009}, is given by $\log E_{\rm iso} = \beta + \gamma\log E_{\rm p}$, where $\gamma$ and $\beta$ are the slope and intercept parameters and the rest-frame peak energy $E_{\rm p}$ of a GRB source is in units of keV and related to the observed peak energy $E_{\rm p, obs}$ by $E_{\rm p} = (1+z)E_{\rm p, obs}$. By simultaneously determining both Amati correlation and cosmological parameters from J220 GRB data one can verify that J220 GRB data are potentially standardizable through the Amati correlation if the resulting Amati correlation parameter values are independent of cosmological model.

The A118 sample, listed in table 7 of Ref.\ \cite{Khadkaetal_2021b}, consists of 118 long GRBs and ranges from $0.3399$ to $8.2$.

\item[]{$H(z)\ +\ BAO\ data$.} Here we use 32 $H(z)$ and 12 BAO measurements listed in Tables 1 and 2 of Ref.\ \cite{CaoRatra2023}, spanning the redshift ranges $0.07 \leq z \leq 1.965$ and $0.122 \leq z \leq 2.334$, respectively.

\end{itemize}

\section{Data Analysis Methodology}
\label{sec:analysis}

The natural log of the GRB data likelihood function is
\begin{equation}
\label{eq:LF_s1}
    \ln\mathcal{L}_{\rm GRB}= -\frac{1}{2}\Bigg[\chi^2_{\rm GRB}+\sum^{N}_{i=1}\ln\left(2\pi\sigma^2_{\mathrm{tot},i}\right)\Bigg],
\end{equation}
where
\begin{equation}
\label{eq:chi2_s1}
    \chi^2_{\rm GRB} = \sum^{N}_{i=1}\bigg[\frac{(\log E_{\mathrm{iso},i} - \beta - \gamma\log E_{\mathrm{p},i})^2}{\sigma^2_{\mathrm{tot},i}}\bigg]
\end{equation}
with total uncertainty $\sigma_{\mathrm{tot},i}$ given by
\begin{equation}
\label{eq:sigma_s2}
\sigma^2_{\mathrm{tot},i}=\sigma_{\mathrm{int}}^2+\sigma_{\log E_{\mathrm{iso},i}}^2+\gamma^2\sigma_{\log E_{\mathrm{p},i}}^2,
\end{equation}
where $\sigma_{\mathrm{int}}$ is the intrinsic scatter parameter for GRB data, which also accounts for unknown systematic uncertainties, \cite{DAgostini2005}.

The likelihood functions for $H(z)$ and BAO data are described in Ref.\ \cite{CaoRatra2023}, which also provides the $H(z)$ + BAO data results shown here.

\begin{table}[htbp]
\centering
% \resizebox{\columnwidth}{!}{%
\begin{threeparttable}
\caption{Flat (uniform) priors of the constrained parameters.}
\label{tab:priors}
\begin{tabular}{lcc}
\toprule\toprule
Parameter & & Prior\\
\midrule
 & Cosmological Parameters & \\
\midrule
$H_0$\,\tnote{a} &  & [None, None]\\
\obhs\,\tnote{b} &  & [0, 1]\\
\ochs\,\tnote{b} &  & [0, 1]\\
\ok &  & [$-2$, 2]\\
$\alpha$ &  & [0, 10]\\
\wx &  & [$-5$, 0.33]\\
\om\,\tnote{c} &  & [0.051314766115, 1]\\
% \midrule
\\
 & Amati Correlation Parameters & \\
% \midrule
$\beta$ &  & [0, 300]\\
$\gamma$ &  & [0, 5]\\
$\sigma_{\mathrm{int}}$ &  & [0, 5]\\
\bottomrule\bottomrule
\end{tabular}
\begin{tablenotes}
\item [a] \hunit. In the GRB cases, $H_0=70$ \hunit.
\item [b] $H(z)$ + BAO. In the GRB cases $\Omega_{b}=0.05$.
\item [c] GRB cases only, to ensure that $\Omega_{c}$ remains positive. For A118, $\Om\in[0,1]$, where $\Omega_{c}$ is the constrained parameter that can be negative due to \textsc{class} version difference.
\end{tablenotes}
\end{threeparttable}%
% }
\end{table}

The parameters we constrain obey flat (uniform) priors, non-zero in the closed intervals listed in Table \ref{tab:priors}. We utilize the {\footnotesize MontePython} Markov chain Monte Carlo (MCMC) code, \cite{Audrenetal2013,Brinckmann2019}, for conducting likelihood analysis of both cosmological and Amati correlation parameters. Subsequent statistical analysis and visualization are facilitated by employing the {\footnotesize GetDist} \textsc{python} package, \cite{Lewis_2019}. For definitions of the Akaike Information Criterion (AIC), Bayesian Information Criterion (BIC), and Deviance Information Criterion (DIC), see Section IV of Ref.\ \cite{CaoRatra2023}). To evaluate model performance, we utilize $\Delta \mathrm{IC}$, which is the difference between the dark energy model IC value and the flat \lcdm\ model baseline IC value. A positive (or negative) $\Delta \mathrm{IC}$ value indicates how much worse (or better) the alternate model fits the dataset in comparison to the baseline flat \lcdm\ model fit. We categorize the strength of evidence against the models based on $\Delta \mathrm{IC}$ values relative to the model with the minimum IC as follows: weak $(0, 2]$, positive $(2, 6]$, strong $(6, 10]$, and very strong $>10$.

\begin{sidewaystable*}
\centering
% \footnotesize
% \fontsize{8.5pt}{10pt}\selectfont
\resizebox{1\columnwidth}{!}{%
% \resizebox*{2.7\columnwidth}{1.8\columnwidth}{%
\begin{threeparttable}
\caption{Unmarginalized best-fitting parameter values for all models from \hiig\ data.}\label{tab:BFP}
\begin{tabular}{lccccccccccccccccc}
\toprule\toprule
Model & Data set & $\Omega_{b}h^2$ & $\Omega_{c}h^2$ & $\Omega_{m0}$ & $\Omega_{k0}$ & $w_{\mathrm{X}}$/$\alpha$\tnote{a} & $H_0$\tnote{b} & $\gamma$ & $\beta$ & $\sigma_{\mathrm{int}}$ & $-2\ln\mathcal{L}_{\mathrm{max}}$ & AIC & BIC & DIC & $\Delta \mathrm{AIC}$ & $\Delta \mathrm{BIC}$ & $\Delta \mathrm{DIC}$ \\
\midrule
Flat \lcdm & $H(z)$ + BAO\tnote{c} & 0.0254 & 0.1200 & 0.297 & -- & -- & 70.12 & -- & -- & -- & 30.56 & 36.56 & 41.91 & 37.32 & 0.00 & 0.00 & 0.00\\%36.5576, 41.91016890175479, 37.318529778549525
 & J220\tnote{d} & -- & -- & 1.000 & -- & -- & -- & 1.379 & 49.17 & 0.382 & 232.58 & 238.58 & 248.76 & 242.21 & 0.00 & 0.00 & 0.00\\%238.578, 248.75888263905708, 242.2058469441781
 & A118\tnote{c,d} & -- & 0.2768 & 0.616 & -- & -- & -- & 1.166 & 49.92 & 0.382 & 118.63 & 126.63 & 137.71 & 125.95 & 0.00 & 0.00 & 0.00\\[6pt]%126.6294, 137.71213849786267, 125.95055999632672
Nonflat \lcdm & $H(z)$ + BAO\tnote{c} & 0.0269 & 0.1128 & 0.289 & 0.041 & -- & 69.61 & -- & -- & -- & 30.34 & 38.34 & 45.48 & 38.80 & 1.78 & 3.56 & 1.48\\%38.3384, 45.475158535673046, 38.79768870799054
 & J220\tnote{d} & -- & -- & 0.999 & 0.432 & -- & -- & 1.394 & 49.12 & 0.384 & 232.32 & 240.32 & 253.90 & 243.15 & 1.74 & 5.14 & 0.94\\%240.322, 253.89651018540945, 243.14754481237412
 & A118\tnote{c,d} & -- & 0.4633 & 0.997 & 1.553 & -- & -- & 1.177 & 49.71 & 0.380 & 117.47 & 127.47 & 141.32 & 126.29 & 0.84 & 3.61 & 0.34\\[6pt]%127.4658, 141.31922312232834, 126.28595007424988
Flat XCDM & $H(z)$ + BAO\tnote{c} & 0.0318 & 0.0938 & 0.283 & -- & $-0.734$ & 66.67 & -- & -- & -- & 26.58 & 34.58 & 41.71 & 34.83 & $-1.98$ & $-0.20$ & $-2.49$\\%34.575, 41.71175853567304, 34.826071644238766
 & J220\tnote{d} & -- & -- & 0.058 & -- & 0.130 & -- & 1.360 & 49.15 & 0.384 & 231.92 & 239.92 & 253.49 & 242.55 & 1.34 & 4.74 & 0.35\\%239.92, 253.49451018540944, 242.55493611035635
 & A118\tnote{c,d} & -- & $-0.0223$ & 0.006 & -- & $-0.203$ & -- & 1.186 & 49.87 & 0.383 & 118.07 & 128.07 & 141.92 & 126.87 & 1.44 & 4.21 & 0.92\\[6pt]%128.07, 141.92302312232832, 126.86662214989649
Nonflat XCDM & $H(z)$ + BAO\tnote{c} & 0.0305 & 0.0998 & 0.293 & $-0.084$ & $-0.703$ & 66.79 & -- & -- & -- & 26.00 & 36.00 & 44.92 & 36.11 & $-0.56$ & 3.01 & $-1.21$\\%35.9978, 44.91874816959131, 36.11102222322448
 & J220\tnote{d} & -- & -- & 0.110 & $-0.265$ & 0.080 & -- & 1.349 & 49.15 & 0.387 & 231.87 & 241.87 & 258.84 & 244.29 & 3.30 & 10.08 & 2.09\\%241.874, 258.8421377317618, 244.29236600808068
 & A118\tnote{c,d} & -- & 0.4579 & 0.986 & 1.260 & $-1.127$ & -- & 1.179 & 49.71 & 0.383 & 117.50 & 129.50 & 146.12 & 126.97 & 2.87 & 8.41 & 1.02\\[6pt]%129.4992, 146.12330774679398, 126.97333446532176
Flat \pcdm & $H(z)$ + BAO\tnote{c} & 0.0336 & 0.0866 & 0.271 & -- & 1.157 & 66.80 & -- & -- & -- & 26.50 & 34.50 & 41.64 & 34.15 & $-2.05$ & $-0.27$ & $-3.17$\\%34.504599999999996, 41.64135853567304, 34.150297701330516
 & J220\tnote{d} & -- & -- & 0.998 & -- & 1.047 & -- & 1.384 & 49.16 & 0.384 & 232.58 & 240.58 & 254.15 & 242.33 & 2.00 & 5.39 & 0.13\\%240.578, 254.15251018540945, 242.33106596915545
 & A118\tnote{c,d} & -- & 0.1226 & 0.301 & -- & 9.805 & -- & 1.173 & 49.89 & 0.383 & 118.24 & 128.24 & 142.10 & 125.51 & 1.62 & 4.39 & $-0.44$\\[6pt]%128.2448, 142.09822312232833, 125.51414811591812
Nonflat \pcdm & $H(z)$ + BAO\tnote{c} & 0.0337 & 0.0894 & 0.275 & $-0.074$ & 1.393 & 67.16 & -- & -- & -- & 25.92 & 35.92 & 44.84 & 35.29 & $-0.64$ & 2.93 & $-2.03$\\%35.921800000000005, 44.8427481695913, 35.2914687221718 D
 & J220\tnote{d} & -- & -- & 0.990 & $-0.356$ & 9.498 & -- & 1.363 & 49.16 & 0.385 & 232.21 & 242.21 & 259.17 & 241.62 & 3.63 & 10.42 & $-0.59$\\%242.206, 259.1741377317618, 241.61750044549987
 & A118\tnote{c,d} & -- & 0.2763 & 0.615 & 0.383 & 6.632 & -- & 1.180 & 49.87 & 0.381 & 118.00 & 130.00 & 146.62 & 126.28 & 3.37 & 8.91 & 0.33\\%129.995, 146.619107746794, 126.27807861071165
\bottomrule\bottomrule
\end{tabular}
%}
\begin{tablenotes}
\item [a] \wx\ corresponds to flat/nonflat XCDM and $\alpha$ corresponds to flat/nonflat \pcdm.
\item [b] \hunit.
\item [c] From Ref.\ \cite{CaoRatra2023}.
\item [d] $\Omega_b=0.05$ and $H_0=70$ \hunit.
\end{tablenotes}
\end{threeparttable}%
}
\end{sidewaystable*}

\begin{sidewaystable*}
\centering
% \footnotesize
\resizebox{1\columnwidth}{!}{%
\begin{threeparttable}
\caption{One-dimensional marginalized posterior mean values and uncertainties ($\pm 1\sigma$ error bars or $2\sigma$ limits) of the parameters for all models from various combinations of data.}\label{tab:1d_BFP}
\begin{tabular}{lcccccccccc}
\toprule\toprule
Model & Data set & $\Omega_{b}h^2$ & $\Omega_{c}h^2$ & $\Omega_{m0}$ & $\Omega_{k0}$ & $w_{\mathrm{X}}$/$\alpha$\tnote{a} & $H_0$\tnote{b} & $\gamma$ & $\beta$ & $\sigma_{\mathrm{int}}$\\
\midrule
Flat \lcdm & $H(z)$ + BAO\tnote{c} & $0.0260\pm0.0040$ & $0.1212^{+0.0091}_{-0.0101}$ & $0.297^{+0.015}_{-0.017}$ & -- & -- & $70.49\pm2.74$ & -- & -- & -- \\
 & J220\tnote{d} & -- & -- & $>0.542$ & -- & -- & -- & $1.397\pm0.066$ & $49.18\pm0.17$ & $0.390^{+0.021}_{-0.024}$ \\%
 & A118\tnote{c,d} & -- & -- & $0.598^{+0.292}_{-0.226}$ & -- & -- & -- & $1.171\pm0.087$ & $49.93\pm0.25$ & $0.393^{+0.027}_{-0.032}$ \\[6pt]%$>0.215$
% \\
Nonflat \lcdm & $H(z)$ + BAO\tnote{c} & $0.0275^{+0.0046}_{-0.0051}$ & $0.1131^{+0.0180}_{-0.0181}$ & $0.289\pm0.023$ & $0.047^{+0.082}_{-0.089}$ & -- & $69.81\pm2.80$ & -- & -- & -- \\
 & J220\tnote{d} & -- & -- & $>0.569$ & $-0.612^{+0.610}_{-0.871}$ & -- & -- & $1.408\pm0.068$ & $49.12\pm0.19$ & $0.390^{+0.021}_{-0.024}$ \\%
 & A118\tnote{c,d} & -- & -- & $>0.267$ & $0.789^{+0.664}_{-0.775}$ & -- & -- & $1.186\pm0.089$ & $49.82\pm0.26$ & $0.392^{+0.026}_{-0.032}$ \\[6pt]
% \\
Flat XCDM & $H(z)$ + BAO\tnote{c} & $0.0308^{+0.0053}_{-0.0046}$ & $0.0978^{+0.0184}_{-0.0164}$ & $0.285\pm0.019$ & -- & $-0.776^{+0.130}_{-0.103}$ & $67.18\pm3.18$ & -- & -- & -- \\
 & J220\tnote{d} & -- & -- & $>0.289$ & -- & $<0.086$ & -- & $1.392\pm0.067$ & $49.19\pm0.18$ & $0.390^{+0.021}_{-0.024}$ \\%
 & A118\tnote{c,d} & -- & -- & $0.557^{+0.277}_{-0.274}$ & -- & $-2.521^{+2.330}_{-2.370}$ & -- & $1.167\pm0.088$ & $50.01^{+0.27}_{-0.31}$ & $0.393^{+0.026}_{-0.032}$ \\[6pt]%$>0.159$
% \\
Nonflat XCDM & $H(z)$ + BAO\tnote{c} & $0.0303^{+0.0054}_{-0.0048}$ & $0.1021\pm0.0193$ & $0.292\pm0.024$ & $-0.054\pm0.103$ & $-0.757^{+0.135}_{-0.093}$ & $67.33\pm2.96$ & -- & -- & -- \\
 & J220\tnote{d} & -- & -- & $>0.296$ & $0.254^{+0.591}_{-0.758}$ & $-1.883^{+2.033}_{-0.824}$ & -- & $1.395\pm0.073$ & $49.13\pm0.20$ & $0.391^{+0.021}_{-0.024}$ \\%
 & A118\tnote{c,d} & -- & -- & $>0.226$ & $0.690^{+0.512}_{-0.798}$ & $-2.342^{+2.067}_{-1.106}$ & -- & $1.185\pm0.090$ & $49.82\pm0.28$ & $0.392^{+0.026}_{-0.032}$ \\[6pt]
% \\
Flat \pcdm & $H(z)$ + BAO\tnote{c} & $0.0326^{+0.0061}_{-0.0030}$ & $0.0866^{+0.0197}_{-0.0180}$ & $0.272^{+0.024}_{-0.022}$ & -- & $1.271^{+0.507}_{-0.836}$ & $66.19^{+2.89}_{-2.88}$ & -- & -- & -- \\%$1.271^{+1.294}_{-1.228}$ 2$\sigma$
 & J220\tnote{d} & -- & -- & $>0.377$ & -- & $>4.007$\tnote{e} & -- & $1.397\pm0.066$ & $49.17\pm0.17$ & $0.390^{+0.020}_{-0.024}$ \\%
 & A118\tnote{c,d} & -- & -- & $0.535^{+0.293}_{-0.287}$ & -- & -- & -- & $1.171\pm0.088$ & $49.88\pm0.24$ & $0.392^{+0.026}_{-0.032}$ \\[6pt]%$>0.124$
% \\
Nonflat \pcdm & $H(z)$ + BAO\tnote{c} & $0.0325^{+0.0064}_{-0.0029}$ & $0.0881^{+0.0199}_{-0.0201}$ & $0.275\pm0.025$ & $-0.052^{+0.093}_{-0.087}$ & $1.427^{+0.572}_{-0.830}$ & $66.24\pm2.88$ & -- & -- & -- \\%$1.427^{+1.365}_{-1.317}$
 & J220\tnote{d} & -- & -- & $>0.412$ & $-0.269^{+0.305}_{-0.275}$ & $>0.462$\tnote{e} & -- & $1.381\pm0.068$ & $49.18\pm0.17$ & $0.390^{+0.020}_{-0.023}$ \\%
 & A118\tnote{c,d} & -- & -- & $0.516^{+0.215}_{-0.288}$ & $0.064^{+0.293}_{-0.282}$ & $5.209^{+3.855}_{-2.462}$ & -- & $1.174\pm0.089$ & $49.88\pm0.24$ & $0.392^{+0.026}_{-0.032}$ \\%$0.516^{+0.452}_{-0.390}$
\bottomrule\bottomrule
\end{tabular}
%}
\begin{tablenotes}
\item [a] \wx\ corresponds to flat/nonflat XCDM and $\alpha$ corresponds to flat/nonflat \pcdm.
\item [b] \hunit.
\item [c] From Ref.\ \cite{CaoRatra2023}.
\item [d] $\Omega_b=0.05$ and $H_0=70$ \hunit.
\item [e] This is the 1$\sigma$ limit. The 2$\sigma$ limit is set by the prior and not shown here.
\end{tablenotes}
\end{threeparttable}%
}
\end{sidewaystable*}

\section{Results}
\label{sec:results}

\begin{figure*}[htbp]
\centering
 \subfloat[]{%
    \includegraphics[width=0.45\textwidth,height=0.35\textwidth]{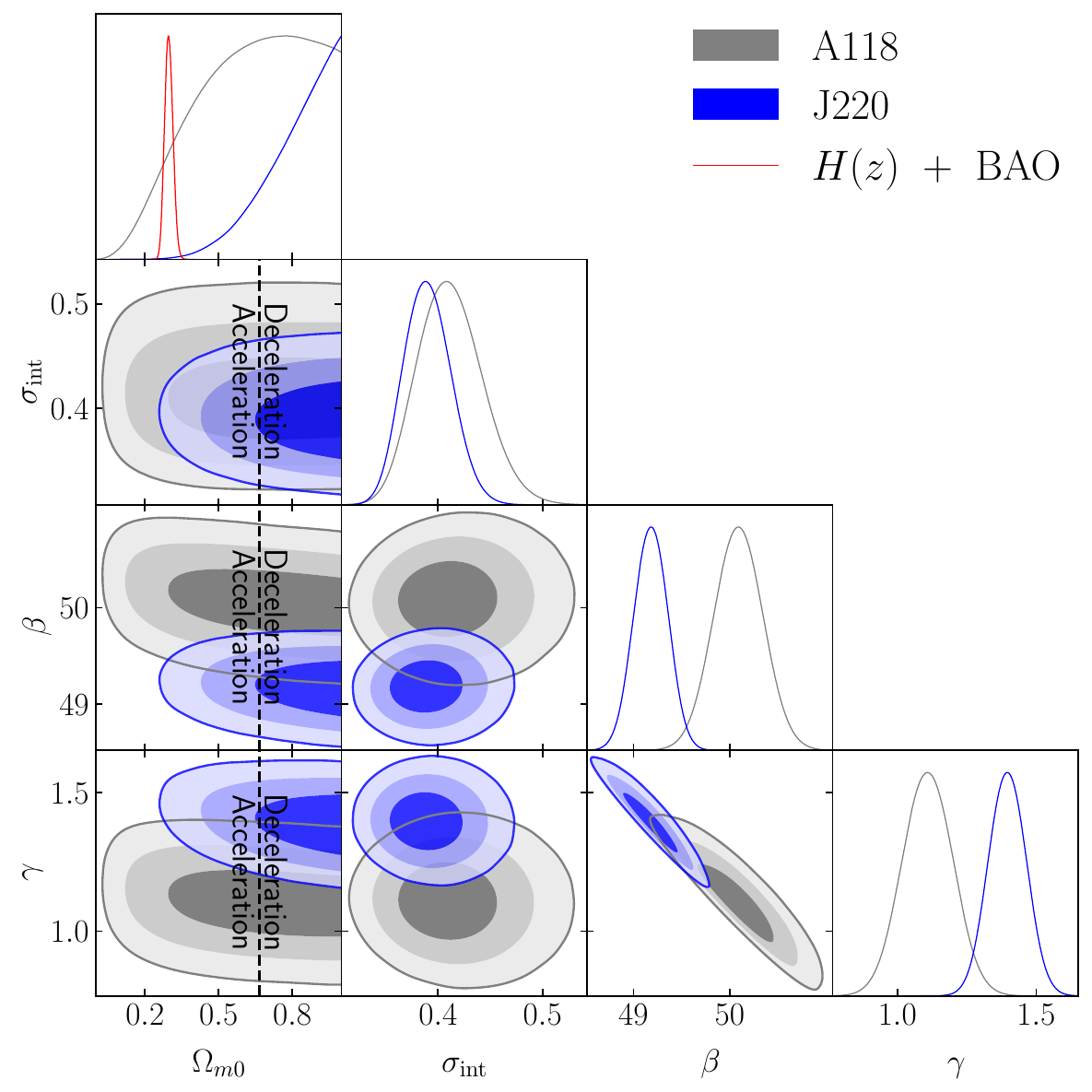}}
 \subfloat[]{%
    \includegraphics[width=0.45\textwidth,height=0.35\textwidth]{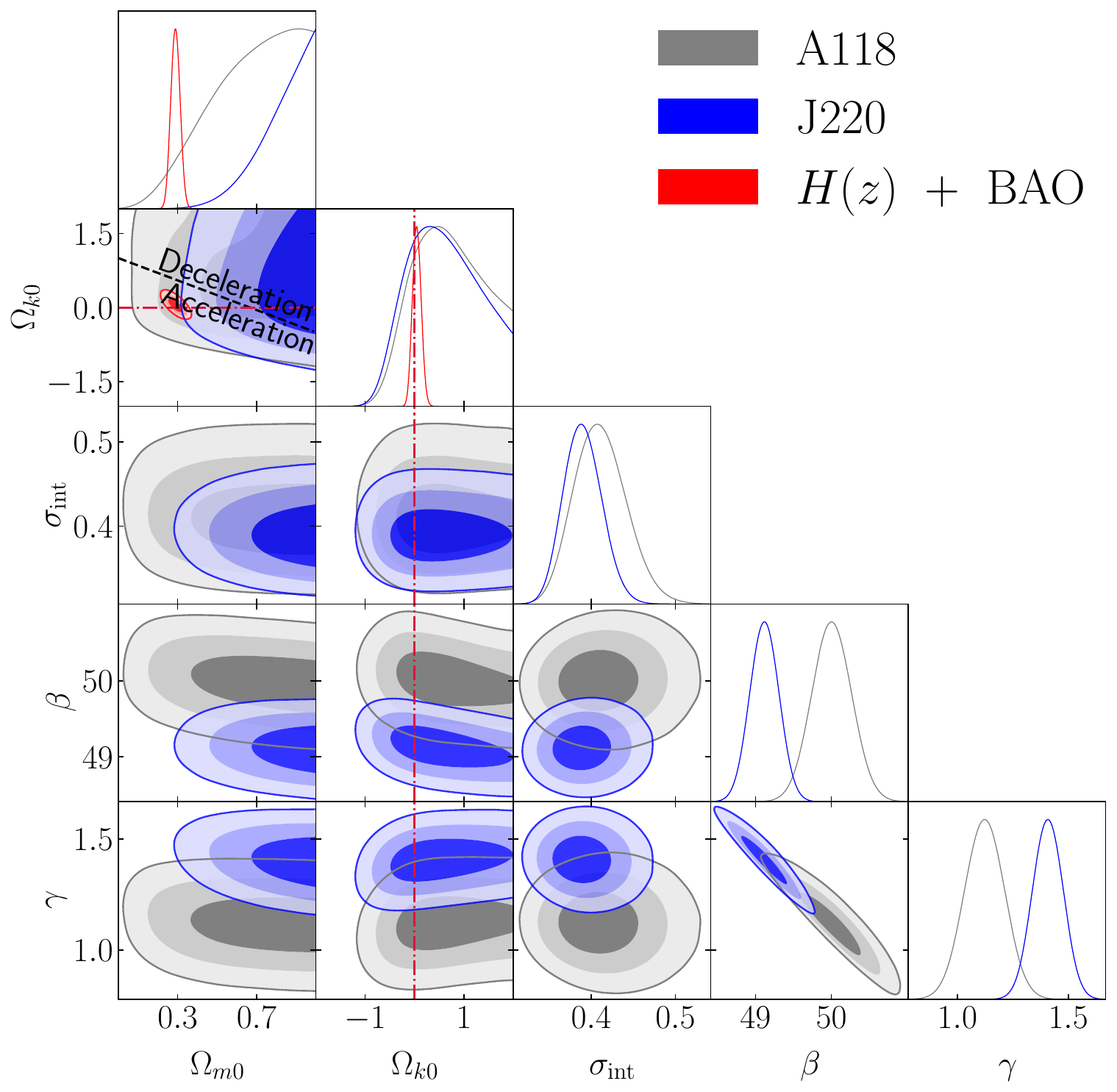}}\\
 \subfloat[]{%
    \includegraphics[width=0.45\textwidth,height=0.35\textwidth]{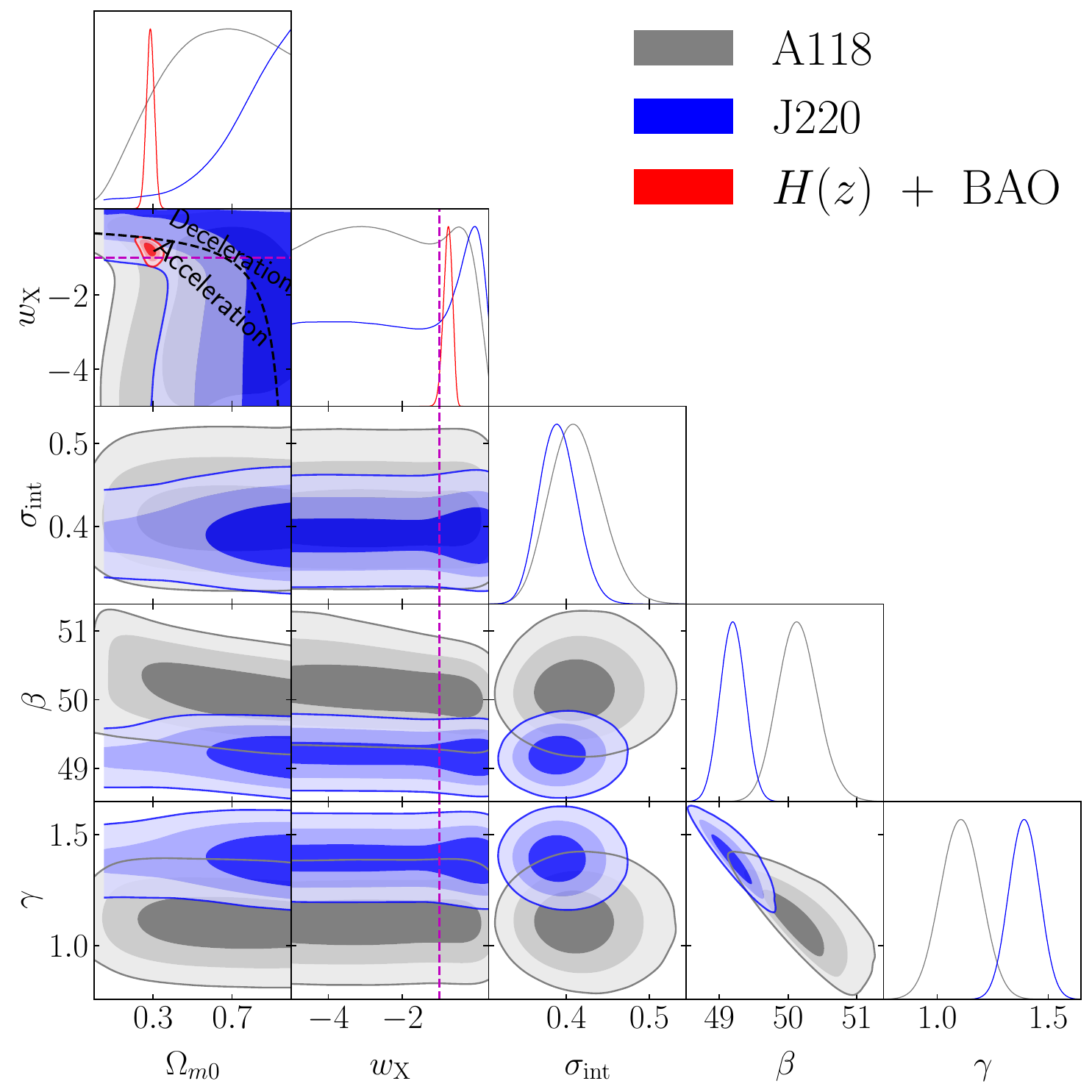}}
 \subfloat[]{%
    \includegraphics[width=0.45\textwidth,height=0.35\textwidth]{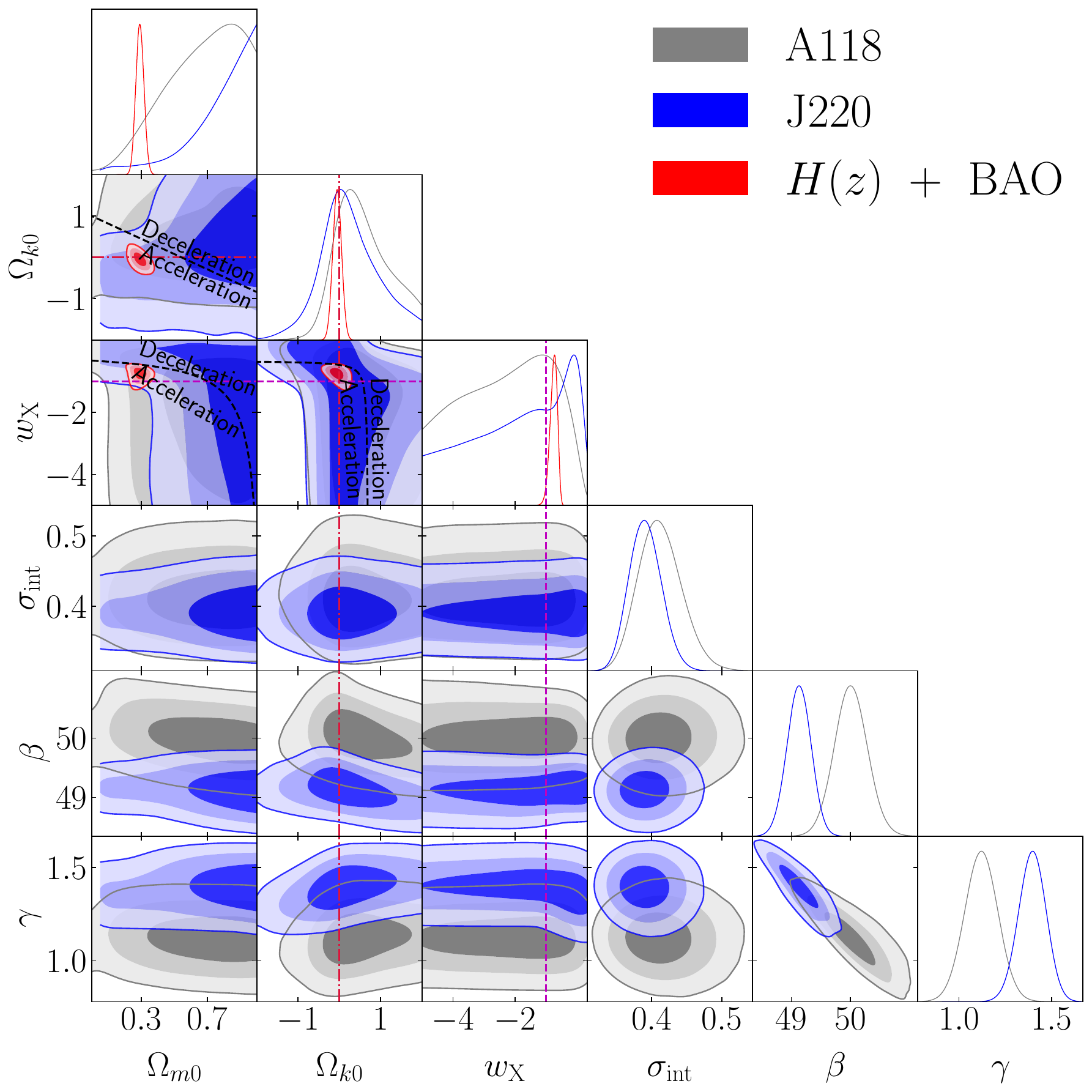}}\\
 \subfloat[]{%
    \includegraphics[width=0.45\textwidth,height=0.35\textwidth]{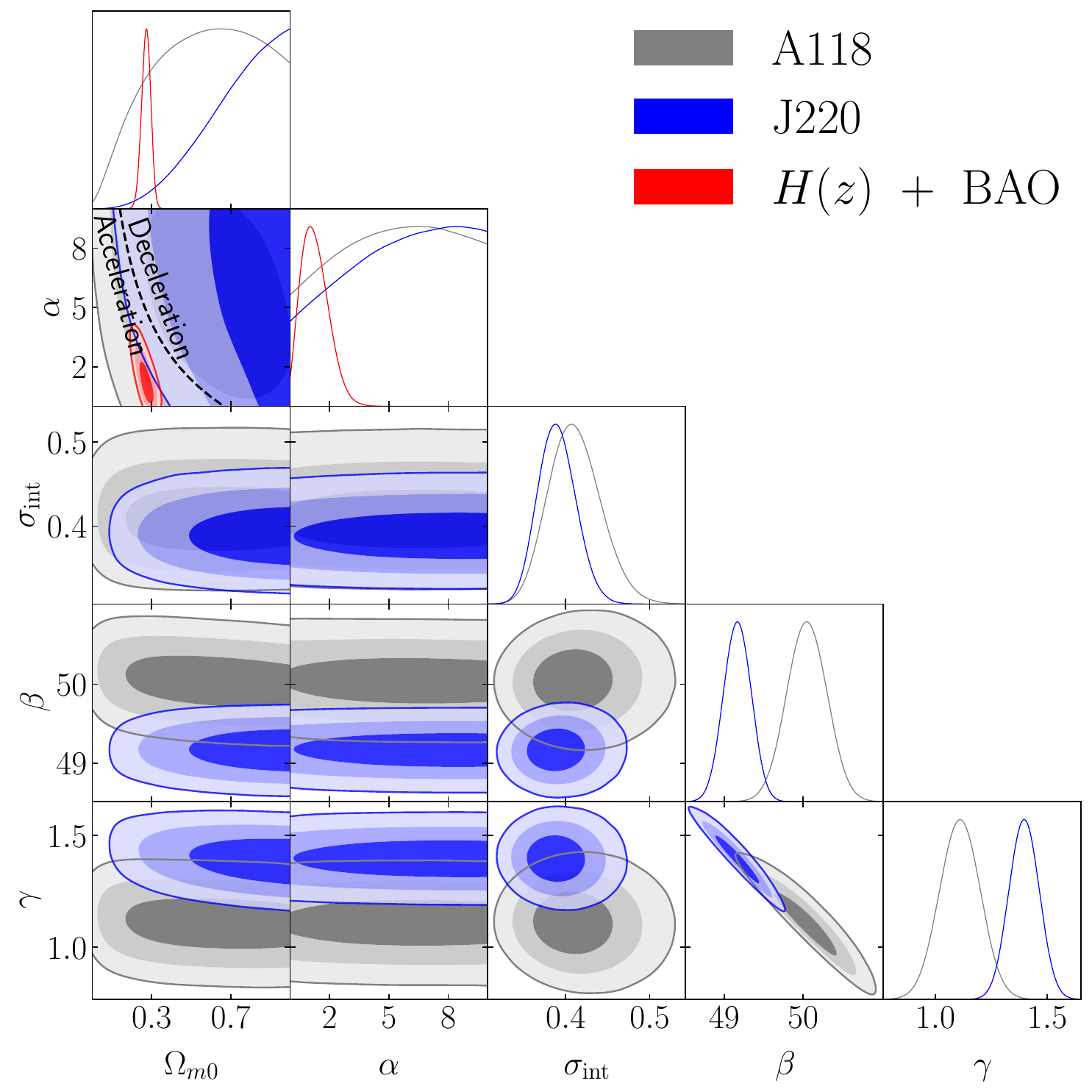}}
 \subfloat[]{%
    \includegraphics[width=0.45\textwidth,height=0.35\textwidth]{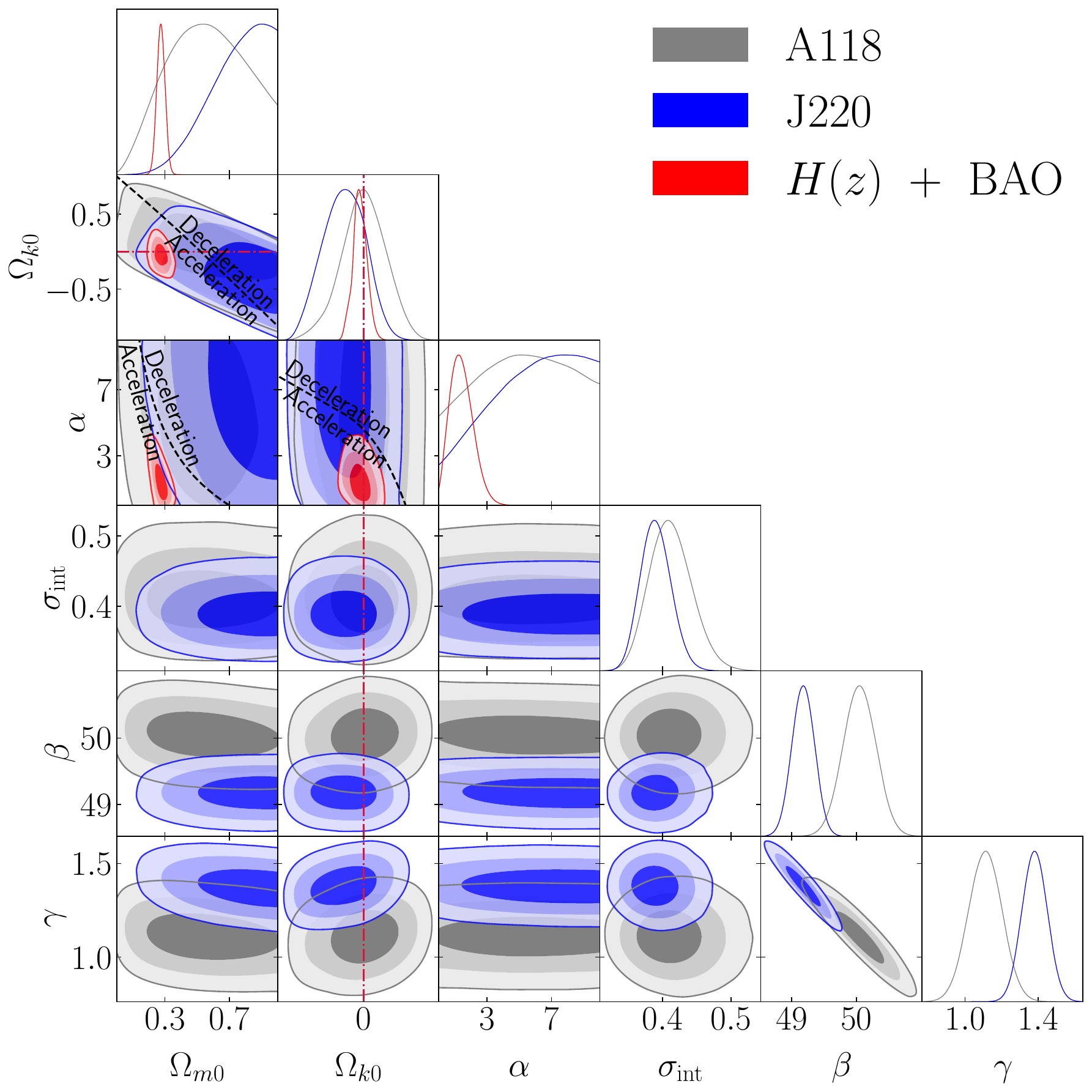}}\\
\caption{One-dimensional likelihoods and 1$\sigma$, 2$\sigma$, and 3$\sigma$ two-dimensional likelihood confidence contours from GRB A118 (gray), J220 (blue), and $H(z)$ + BAO (red) data for six different models, with \lcdm, XCDM, and \pcdm\ in the top, middle, and bottom rows, and flat (nonflat) models in the left (right) column. The black dashed zero-acceleration lines, computed for the third cosmological parameter set to the $H(z)$ + BAO data best-fitting values listed in Table \ref{tab:BFP} in panels (d) and (f), divide the parameter space into regions associated with currently-accelerating (below or below left) and currently-decelerating (above or above right) cosmological expansion. The crimson dash-dot lines represent flat hypersurfaces, with closed spatial hypersurfaces either below or to the left. The magenta lines represent $w_{\rm X}=-1$, i.e.\ flat or nonflat \lcdm\ models. The $\alpha = 0$ axes correspond to flat and nonflat \lcdm\ models in panels (e) and (f), respectively.}
\label{fig1}
\vspace{-50pt}
\end{figure*}

\begin{figure*}
\centering
 \subfloat[]{%
    \includegraphics[width=0.4\textwidth,height=0.35\textwidth]{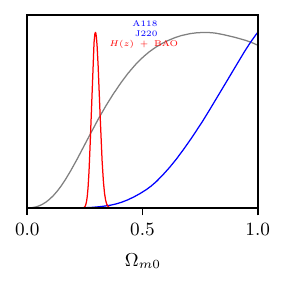}}
 \subfloat[]{%
    \includegraphics[width=0.4\textwidth,height=0.35\textwidth]{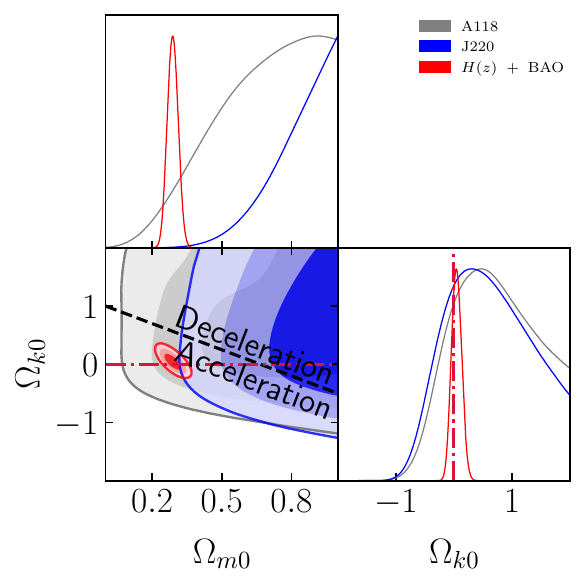}}\\
 \subfloat[]{%
    \includegraphics[width=0.4\textwidth,height=0.35\textwidth]{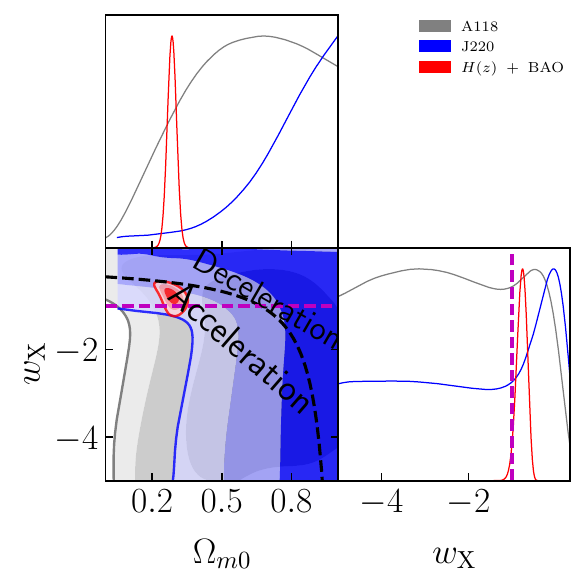}}
 \subfloat[]{%
    \includegraphics[width=0.4\textwidth,height=0.35\textwidth]{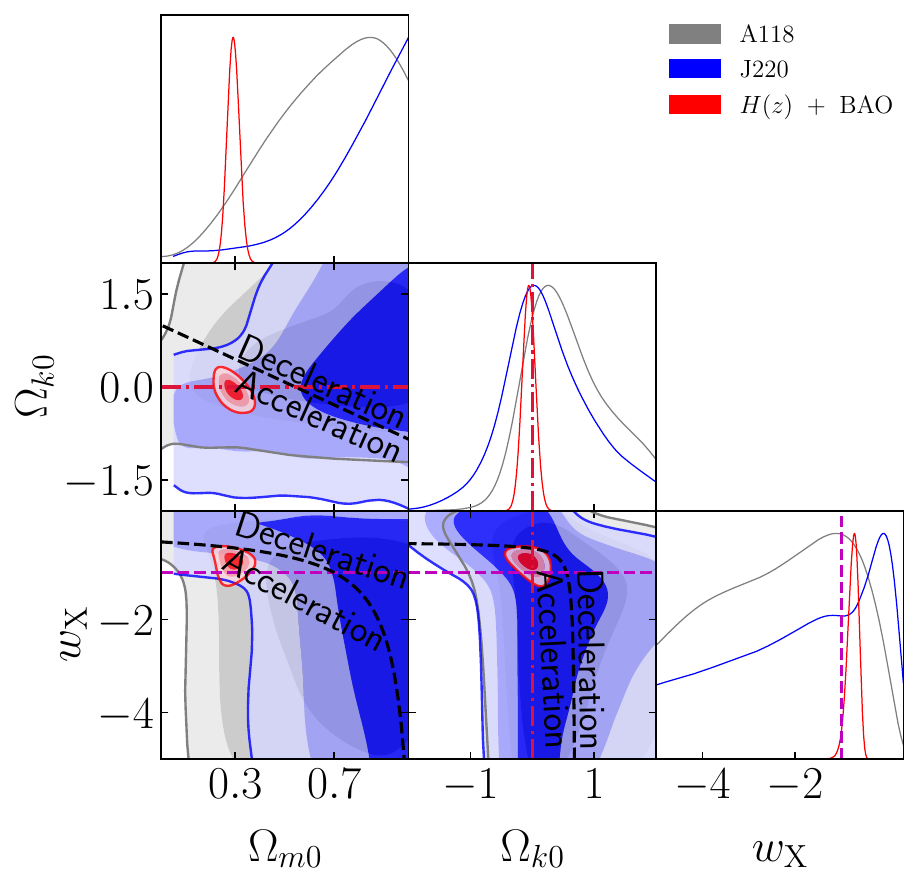}}\\
 \subfloat[]{%
    \includegraphics[width=0.4\textwidth,height=0.35\textwidth]{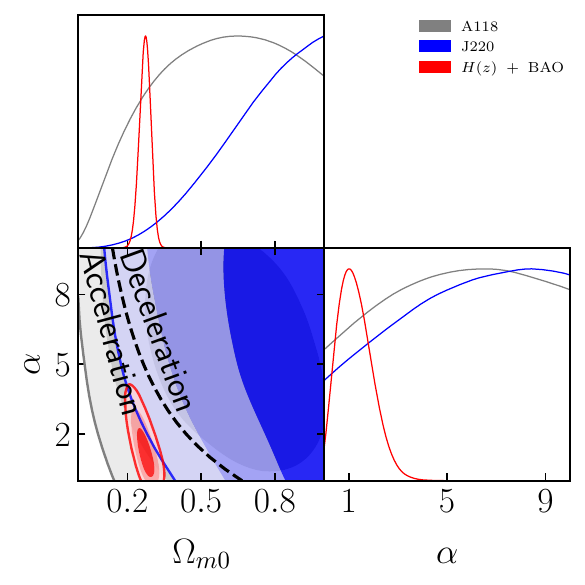}}
 \subfloat[]{%
    \includegraphics[width=0.4\textwidth,height=0.35\textwidth]{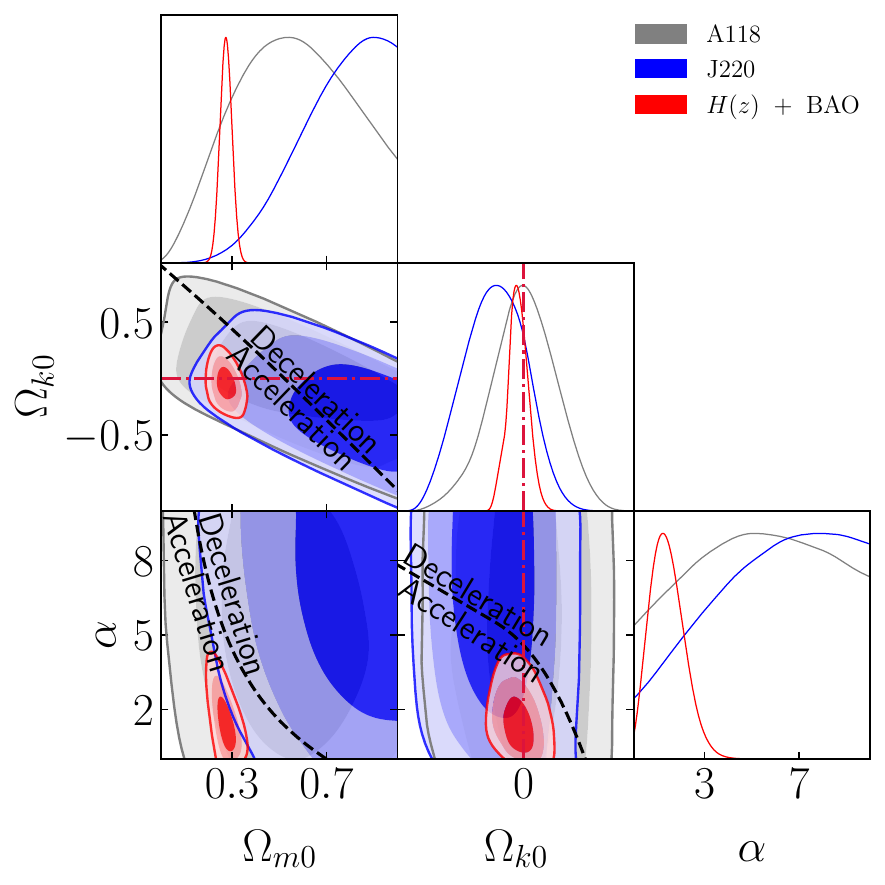}}\\
\caption{Same as Figure \ref{fig1}, but for cosmological parameters only.}
\label{fig2}
\end{figure*}

The constraints of A118 and $H(z)$ + BAO data are taken from Ref.\ \cite{CaoRatra2023} and listed in Tables \ref{tab:BFP} and \ref{tab:1d_BFP}. The values of the unmarginalized best-fitting parameters, maximum likelihood $\mathcal{L}_{\rm max}$, AIC, BIC, DIC, $\Delta \mathrm{AIC}$, $\Delta \mathrm{BIC}$, and $\Delta \mathrm{DIC}$ for all models and datasets are presented in Table \ref{tab:BFP}. Additionally, the one-dimensional marginalized posterior mean parameter values and their uncertainties ($\pm 1\sigma$ error bars or 1 or 2$\sigma$ limits) for all models and datasets can be found in Table \ref{tab:1d_BFP}. Figs.\ \ref{fig1} and \ref{fig2} show the posterior distributions and contours of all parameters and of just cosmological parameters, respectively, for the six cosmological models under study, where the J220 and $H(z)$ + BAO results are shown in blue and red, respectively.

We have found that only for the XCDM parametrizations are the constraints on cosmological parameters from J220 data consistent with those from $H(z)$ + BAO data. Specifically, for flat XCDM the \om\ and \wx\ constraints are $\Omega_{m0} >0.289$ [$=0.285\pm0.019$] and $w_X <0.086$ [$=-0.776^{+0.130}_{-0.103}$] for J220 [$H(z)$ + BAO] data, respectively; and for nonflat XCDM the \om, \wx, and \ok\ constraints are $\Omega_{m0} >0.296$ [$=0.292\pm0.024$], $w_X = -0.054\pm0.103$ [$-1.883^{+2.033}_{-0.824}$], and $\Omega_{k0} = -0.757^{+0.135}_{-0.093}$ [$0.254^{+0.591}_{-0.758}$] for J220 [$H(z)$ + BAO] data, respectively. However, for other models the constraints on \om\ exhibit tensions exceeding $2\sigma$, while the J220 and $H(z)$ + BAO data constraints on \ok\ and $\alpha$ are consistent within $2\sigma$.

Given that, more so than not, the J220 constraints on \om\ are not consistent with those from better-established $H(z)$ + BAO data, these J220 GRB data cannot be jointly used with better-established data for the purpose of constraining cosmological parameters.

It is noteworthy that the J220 and $H(z)$ + BAO data constraints on \ok\ for all three nonflat models are consistent within $1\sigma$. Furthermore, all J220 cases favor closed hypersurfaces, however only the nonflat \lcdm\ model exhibits a slight bias (just $1.003\sigma$) away from flat geometry.

The constraints on the Amati correlation slope parameter $\gamma$ range from a low of $1.381\pm0.068$ (nonflat \pcdm) to a high of $1.408\pm0.068$ (nonflat \lcdm), with maximum difference of $0.28\sigma$. The constraints on the Amati correlation intercept parameter $\beta$ range from a low of $49.12\pm0.19$ (nonflat \lcdm) to a high of $49.19\pm0.18$ (flat XCDM), with maximum difference of $0.27\sigma$. Therefore we conclude that J220 GRB data are standardizble through the Amati correlation. 

The constraints on the intrinsic scatter parameter $\sigma_{\rm int}$ range from a low of $0.390^{+0.020}_{-0.024}$ (flat \pcdm) to a high of $0.391^{+0.021}_{-0.024}$ (nonflat XCDM), with maximum difference of $0.03\sigma$. The intrinsic scatter for the J220 sample in this work ($\sigma_{\rm int}\sim0.39$) are consistent with the A118 sample from Ref. \cite{Khadkaetal_2021b}, which may be due to the 7 updated GRB data (080916C, 090323, 090328, 090424, 090902B, 091127, 130427A) from the A118 sample with high-quality spectral fitting models than the Band model used in the J220 sample.

For J220 data, based on AIC (BIC), flat \lcdm\ is the most favored model, the evidence against nonflat XCDM and nonflat \pcdm\ is positive (very strong), and the evidence against the rest is weak (positive). Based on the more reliable DIC, nonflat \pcdm\ is the most favored model, the evidence against nonflat XCDM is positive, and the evidence against the rest is weak.

Due to the large error bars, the J220 constraints on cosmological parameters are largely consistent with the A118 constraints, with the largest difference being in \ok\ of nonflat \lcdm\ which is $1.42\sigma$, where the two favor opposing spatial geometry. However, A118 constraints on cosmological parameters are mostly consistent with $H(z)$ + BAO constraints, with the largest difference being in $\alpha$\ of flat \pcdm\ which is slightly less than $1.50\sigma$ and not that significant.

As shown in Table \ref{tab:diff}, the J220 constraints on the Amati correlation parameters, slope $\gamma$ and intercept $\beta$, are largely inconsistent with the A118 constraints. The differences range from a maximum of $2.07\sigma$ for $\gamma$ and $2.51\sigma$ for $\beta$ (both in flat \lcdm) to a maximum of $1.81\sigma$ for $\gamma$ and $2.01\sigma$ for $\beta$ (both in nonflat XCDM). The intrinsic scatter for J220 is slightly smaller than that for A118 across all models, with differences ranging from $0.03\sigma$ to $0.08\sigma$. This may be due to the larger dataset, the updated better-quality data for some of the GRBs, differences in data quality, or a combination of these factors. 

\begin{table*}
\centering
% \resizebox{2\columnwidth}{!}{%
\setlength\tabcolsep{14pt}
\begin{threeparttable}
\caption{The differences between A118 and J220 for a given cosmological model with $1\sigma$ being the quadrature sum of the two corresponding $1\sigma$ error bars.}\label{tab:diff}
\begin{tabular}{lccc}
\toprule
 Model & $\Delta\gamma$ & $\Delta\beta$ & $\Delta\sigma_{\mathrm{int}}$ \\
\midrule
Flat \lcdm & $2.07\sigma$ & $2.51\sigma$ & $0.08\sigma$ \\
Nonflat \lcdm & $1.98\sigma$ & $2.17\sigma$ & $0.05\sigma$ \\
Flat XCDM & $2.03\sigma$ & $2.29\sigma$ & $0.08\sigma$ \\
Nonflat XCDM & $1.81\sigma$ & $2.01\sigma$ & $0.03\sigma$ \\
Flat \pcdm & $2.05\sigma$ & $2.41\sigma$ & $0.05\sigma$ \\
Nonflat \pcdm & $1.85\sigma$ & $2.38\sigma$ & $0.05\sigma$ \\
\bottomrule
\end{tabular}
%}
% \begin{tablenotes}[flushleft]
% \item [a] Only for flat/non-flat \lcdm.

% \end{tablenotes}
\end{threeparttable}%
% }
\end{table*}

\section{Conclusion}
\label{sec:conclusion}

We use updated J220 GRB data based on the compilation of Ref.\ \cite{Jiaetal2022} to simultaneously constrain Amati correlation parameters and cosmological parameters in six spatially flat and nonflat dark energy cosmological models. We have found that these updated 220 GRBs are standardizable through the Amati correlation, because their constraints on both Amati correlation parameters are independent of cosmological model. However, we have also found that their constraints on \om\ are in $>2\sigma$ tension with those from better-established $H(z)$ + BAO data for the four flat and nonflat \lcdm\ and \pcdm\ models and so these J220 GRB data cannot be used to constrain cosmological parameters. The J220 intrinsic scatter constraints $\sigma_{\rm int}\sim0.39$ are fairly consistent with the A118 constraints reported in Ref.\ \cite{Khadkaetal_2021b} (and A118 constraints here), while the A102 constraints are $\sim0.52$ and the joint A220 constraints are between the two at $\sim0.46$, indicating that A118 is of significantly better quality than A102. However, despite both A118 and A102 providing \om\ constraints consistent with those from $H(z)$ + BAO, their combined A220 constraints are inconsistent with those from $H(z)$ + BAO, which suggests that A102 is not fully compatible with A118 for constraining cosmological parameters. While J220 exhibits a similar level of intrinsic scatter to A118, it does not yield \om\ constraints consistent with $H(z)$ + BAO. Consequently, the A118 GRB dataset \citep{Khadkaetal_2021b, LuongoMuccino2021, CaoKhadkaRatra2021, CaoDainottiRatra2022, Liuetal2022} remains the most reliable and largest dataset for cosmological purposes.

% \appendix
% \section{Some title}
% Please always give a title also for appendices.

\acknowledgments

The computations for this project were performed on the Beocat Research Cluster at Kansas State University, which is funded in part by NSF grants CNS-1006860, EPS-1006860, EPS-0919443, ACI-1440548, CHE-1726332, and NIH P20GM113109.

% \paragraph{Note added.} This is also a good position for notes added after the paper has been written.

% Bibliography

%% [A] Recommended: using JHEP.bst file
\bibliographystyle{JHEP}
\bibliography{biblio.bib}

%% or
%% [B] Manual formatting (see below)
%% (i) We suggest to always provide author, title and journal data or doi:
%% in short all the informations that clearly identify a document.
%% (ii) please avoid comments such as "For a review'', "For some examples",
%% "and references therein" or move them in the text. In general, please leave only references in the bibliography and move all
%% accessory text in footnotes.
%% (iii) Also, please have only one work for each \bibitem.

\end{document}